\documentclass[11pt]{article}
\usepackage{graphicx}

\begin{document}
%


\begin{center}
{\large \bf Phase Transitions in the Early Universe} \vskip.5cm

W-Y. Pauchy Hwang\footnote{Email: wyhwang@phys.ntu.edu.tw}$^{,1}$ and
Sang Pyo Kim$^2$ \\
{\em $^1$Asia Pacific Organization for Cosmology and Particle Astrophysics,
\\Center for Theoretical Sciences, Institute of Astrophysics,
\\and Department of Physics, National
Taiwan University, Taipei 106, Taiwan \\
$^2$Department of Physics, Kunsan National University, Kunsan 573-701, Korea, and\\
Asia Pacific Center for Theoretic Physics, Pohang 790-784, Korea  } \vskip.2cm


{\small(25 August 2011; Revised: 13 January 2017)}
\end{center}

\begin{abstract}
The physics of the 20th Century is governed by two
pillars, Einstein's relativity principle and the
quantum principle. At the beginning of the 21st
Century, it becomes clear that there exist the
smallest units of matter, such as electrons,
neutrinos, and quarks; their behaviors are
described by the Standard Model.

It was believed that the temperature of the early
Universe was once $300\, GeV$, or higher, at
$10^{-11} sec$, and then going through the
electroweak phase transition. But the mass phase
transition happens in the purely imaginary 
temperature. Later on, its temperature was 
$150\, MeV$ at $3.3 \times 10^{-5} sec$, and 
then going through the "QCD cosmological phase 
transition". We attempt to use the Standard 
Model, a completely dimensionless theory 
apart from the negative "ignition" term, 
to conclude that the EW or mass phase 
transition {\it does not exist}.

On the front of QCD cosmological phase transition,
the intriguing question about the latent heat
(energy) is discussed and its role is speculated.

\medskip

{\parindent=0truept PACS Indices: 05.30.Ch (Quantum ensemble theory);
98.80.Bp (Origin and formation of the Universe); 98.80.-k (Cosmology).}

{\parindent=0truept Keywords: Origin and formation of the
Universe; Cosmology; Observational cosmology.}
\end{abstract}

\bigskip

\section{Introduction}

Cosmological phase transitions, either electroweak or QCD,
play the intriguing and magic roles in our Universe (or,
our World). Mathematics-wise, it is based on the {\it
same} lagrangian but, to us inside, the worlds before and
after are completely different (physics-wise).

Lately, there were several progresses in the development of the
standard model (i.e., the ultimate theory of elementary
particles), culminating the final form of the Standard Model
\cite{Hwang417}. Even though the ultimate final form of the Standard
Model might still vary, the current form of the Standard Model
\cite{Hwang417} is consistent and perhaps also complete, since
it is well behaved as the relevant distance goes to zero;
also, and it is free from Landau ghosts, as the
momentum-transfer-squared $Q^2$ goes to infinite.

Apart from the "ignition" term, the Standard Model
\cite{Hwang417} turns out to be completely
dimensionless in the quantum 4-dimensional Minkowski
space-time. In this paper,
we wish to perform some initial analysis of this
remarkable aspect, along the line of phase
transitions.

By "quantum" of "the quantum 4-dimensional Minkowski
space-time", we mean at least that the anti-commuting
entities, which might carry some burden of the quantum
principle, also are our main concerns in this
space-time. The meaning of "the quantum 4-dimensional
Minkowski space-time" is certainly different from
that of "the 4-dimensional Minkowski space-time"
- since the quantum principle also allows a lot
of anti-commuting objects. It is different far
from something derived from the $n-$dimensional
complex number system. In other words, the
domain of our interest is greatly enlarged
by the quantum principle.

Thus, the analysis in this paper could be very different
from what was in our earlier investigations
\cite{Hwang} for the cosmological QCD phase transition.
Namely, we focus on the dimensionless nature of the
Standard Model \cite{Hwang417}, i.e., on the "mass"
phase transition (or the "electroweak" transition in
the old terminology).

It is in fact remarkable to realize that, apart
from some Higgs "ignition" term, the Standard
Model \cite{Hwang417} is a completely dimensionless
theory - it seems like that, above some critical 
temperature $T_c$, it is a completely dimensionless 
theory; only below $T_c$, the term "mass" assumes 
its presence and things become rather messy. We
might call it as "the mass phase transition"; we
can accomplish this phase transition by the
Higgs spontaneous symmetry breaking (SSB). Thus,
we could re-name "the EW phase transition"
as "the mass phase transition".

In a much earlier publication \cite{Hwang},
we show the importance of the latent heat,
basing on the MIT or Friedberg-Lee bag
model. It was also suggested that the
latent "heat" or latent energy released,
of course evolving into different forms
of matter, could be linked with dark
matter - an argument only by a simple 
arithmetic.

In this paper, we characterize the "mass" phase transition
as the spontaneous symmetry breaking (SSB), related to the
"origin of mass" \cite{Origin}. That is, the phase
transition in the purely imaginary temperature 
(as accomplished by the hand of the God), related to 
"mass", and the spontaneous symmetry breaking (SSB), 
related to the origin of mass, amounts to the same thing.

Mathematics-wise, the spontaneous symmetry 
breaking (SSB) in the Standard Model cannot
happen in real temperature - rather, if there
would be the purely imaginary temperature, 
then it would go. Either the electroweak 
phase transition or the mass phase transition
{\it simply does not exist}, if we would take 
the hint of our presentation of the Standard
Model \cite{Hwang417} -- apart from the negative
"ignition" term, it is simply a dimensionless 
theory.  

\bigskip

\section{The Standard Model on the Smallest Units of Matter}

What is the Standard Model on the smallest units of matter?
It is a theory \cite{Hwang417} that describes the
behaviors of the point-like particles such as electrons,
neutrinos, quarks, etc. The {\it existence} of the {\it
smallest} units of matter, that is, these point-like
particles, a finite number of them, is so well-established
that it should be next to the two basic pillars, i.e.,
Einstein's relativity principle and the quantum principle.

These particles are "point-like" since the quantum
principle is there; they would be "points" if Newton's
classic thinking would still be there.

What is the Standard Model of All Centuries?
We believe that the description of the point-like
particles, the smallest units of matter in our Universe,
on the basis of the Einstein's relativity principle and
the quantum principle is so fundamental; and it is so
consistent and (perhaps) so complete, that this Standard
Model would stay there longer than Newton's classic
era (of over four hundred years).

In 1970's, S. Weinberg called such theory "the Standard
Model". Since then, such name sticks.

Back in the 20th Century, we didn't realized that there is
something special for the complex scalar fields in the quantum
4-dimensional Minkowski space-time. A complex scalar field
$\phi(x)$ is in fact born to be self-repulsive, due to
the $\lambda (\phi^\dagger \phi)^2$ interaction with
the positive dimensionless coupling $\lambda$. With a
negative $\lambda$, the system will collapse. Thus, if
alone, {\it such field cannot exist.}

For the two complex scalar fields, the attractive mutual
interaction $-2\lambda (\phi_1^\dagger \phi_2) (\phi_2^\dagger \phi_1)$
would be enough to overcome the self-repulsiveness of the two
individual complex scalar fields. The $\lambda$-interaction,
$\lambda (\phi^\dagger \phi)^2$, in fact re-writes the story
for everything.

The story which one of us put forward above about the
{\it nonexistence} of a single complex scalar field
rather striking - whether it is right or not is still
waiting for a clear-cut mathematical proof. Basically,
$\lambda$ cannot be negative since the system would collapse to
the negative infinity. It cannot be zero since this would be
meta-stable. We suspect $\lambda={1\over 8}$ in our notations, but
we have to admit that it is still lack of rigorous proof.

Among the force fields or gauge fields, most of the gauge fields,
upon spontaneous symmetry breaking (SSB), become massive, including
weak bosons and family gauge bosons, if we assume the standard
wisdom \cite{Hwang417}. That calls for the Standard-Model (SM)
Higgs $\Phi(1,2)$, the purely family Higgs $\Phi(3,1)$, and the
mixed family Higgs $\Phi(3,2)$ - they interact maximally
attractively whenever possible, i.e., between $\Phi(1,2)$ and
$\Phi(3,2)$, and between $\Phi(3,1)$ and $\Phi(3,2)$. Here the
two numbers/labels are referred to $SU_f(3)$ and $SU_L(2)$ -
that is, triplets, doublets, or singlets.

So, if there would be only the SM Higgs $\Phi(1,2)$,
then {\it it cannot exist.} The self-repulsive
interaction $\lambda (\phi^\dagger \phi)^2$ (with
$\lambda > 0$) would make it non-existent. As said
earlier, $\lambda=0$ would be meta-stable while
$\lambda<0$ would collapse. The question is why
it is $\lambda={1\over 8}$ and it is dimensionless,
and thus that should be determined {\it globally}
by the quantum 4-dimensional Minkowski space-time.

The algebra among the three Higgs $\Phi(1,2)$,
$\Phi(3,2)$, and $\Phi(3,1)$ arises {\it only}
when it is in the quantum 4-dimensional Minkowski
space-time. If the space-time differs from the
4-dimensional, the algebra simply doesn't apply.

The next question associated with the Higgs fields is to understand
"the origin of mass" - a question that we have recently
gained some understanding \cite{Origin}. In that \cite{Origin}, we
may set all the mass terms of the various Higgs to identically zero,
except one spontaneous-symmetry-breaking (SSB) ignition term. All
the mass terms are the results of this SSB, when switched on.
Apart from "ignition" term, the theory is completely dimensionless
in the 4-dimensional Minkowski space-time.

The set of the "various" Higgs includes the Standard-Model (SM) Higgs
$\Phi(1,2)$, the mixed family Higgs $\Phi(3,2)$, and the pure family
Higgs $\Phi(3,1)$. The "ignition" could be on the
pure family Higgs $\Phi(3,1)$ \cite{Origin}, and it is clear that
that it may not be on the SM Higgs $\Phi(1,2)$.

To make it clear, as the temperature is high enough,
we are dealing with the completely dimensionless
theory, implied by the Standard Model. All the
complications, including some dimensional couplings
(or, the interactions), is the result of
switching-on the "ignition" term.

We are dealing with the phase transition of
the Standard Model \cite{Hwang417} - before
SSB, it is by a set of "pure" numbers (i.e.,
dimensionless couplings) in the quantum
4-dimensional Minkowski space-time. After
SSB, the real world of the Standard Model
shows up. Though it is completely clear to
us, the fundamental secrets about
this set of pure numbers should eventually
be understood. The hypothetical "mass" 
phase transition as represented by the 
Standard Model \cite{Hwang417} happens on 
the axis of the purely imaginary temperature.

\medskip

\subsection{The Dimensionless Lepton World}

The overall background in our world is the quantum 4-dimensional
Minkowski space-time with the force-field gauge-group structure
$SU_c(3) \times SU_L(2) \times U(1) \times SU_f(3)$ imprinted
at the very beginning. It sees the lepton world, of atomic sizes.
It also sees the quark world, of much smaller nuclear sizes.

It is of importance to recognize that the
$SU_L(2) \times U(1) \times SU_f(3)$ symmetry is realized
in the lepton world, through the proposal
\cite{HwangYan} $((\nu_\tau,\,\tau)_L,\,(\nu_\mu,\,\mu)_L,
\,(\nu_e,\,e)_L)$ $(columns)$ ($\equiv \Psi(3,2)$) as the
$SU_f(3)$ triplet and $SU_L(2)$ doublet. In particular, it
is important to realize the role of neutrino oscillations
- it is the change of a neutrino in one generation (flavor)
into that in another generation; or, we need to have the
coupling similar to $i h \bar \Psi_L(3,2)\times
\Psi_R(3,1) \cdot \Phi(3,2)$, as introduced by Hwang
and Yan \cite{HwangYan}. Then, it is clear that the mixed family Higgs
$\Phi(3,2)$ must be there. The remaining purely family Higgs $\Phi(3,1)$
helps to complete the picture, so that the eight gauge bosons are
massive in the $SU_f(3)$ family gauge theory \cite{Family}.

Usually in an old textbook \cite{Books}, the QCD chapter precedes the one
on Glashow-Weinberg-Salam (GWS) electroweak theory, but we are talking
about the $SU_c(3) \times SU_L(2) \times U(1) \times SU_f(3)$ Minkowski
space-time and what happens in it. The so-called "basic units of motion"
are made up from quarks (of six flavors, of three colors, and of
the two helicities) and leptons (of three generations and of the
two helicities). We use these basic units (of motion) in
writing down the lagrangian, etc. - the starting point of
our formalism(s).

If we look at the basic units of motion as compared to the original
particle, i.e. the electron, the starting basic units are all
"point-like" Dirac particles. Dirac invented the Dirac equation
for the electron eighty years ago and surprisingly enough these
"point-like" Dirac particles are the basic units of the
Standard Model. Thus, one of us (Hwang) called it "Dirac
Similarity Principle" - a salute to Dirac; a triumph to
mathematics. Our world could indeed be described by the
proper mathematics.

For the lepton world or the quark world, the story is fixed if
the so-called "gauge-invariant derivative", i.e. $D_\mu$ in
the kinetic-energy term $-\bar \Psi \gamma_\mu D_\mu \Psi$, is
given for a given basic unit, one on one \cite{Books}.

On the lepton world, we introduce the family triplet,
$(\nu_\tau^R,\,\nu_\mu^R,\,,\nu_e^R)$ (column), under $SU_f(3)$.
We write, for $(\nu_\tau^R,\, \nu_\mu^R,\,\nu_e^R)$,
\begin{equation}
D_\mu = \partial_\mu - i \kappa {\bar\lambda^a\over 2} F_\mu^a.
\end{equation}
and, for the left-handed $SU_f(3)$-triplet and $SU_L(2)$-doublet
$((\nu_\tau^L,\,\tau^L),\, (\nu_\mu^L,\,\mu^L),\, (\nu_e^L,\,e^L))$
(all columns),
\begin{equation}
D_\mu = \partial_\mu - i \kappa {\bar\lambda^a\over 2} F_\mu^a - i g
{\vec \tau\over 2} \cdot \vec A_\mu + i {1\over 2} g' B_\mu.
\end{equation}
The right-handed charged leptons form the triplet $\Psi_R^C(3,1)$ under
$SU_f(3)$, since it were singlets their common factor $\bar\Psi_L(\bar 3,2)
\Psi_R(1,1)\Phi(3,2)$ for the mass terms would involve the cross terms such as
$\mu\to e$.

The neutrino mass term assumes a new form \cite{HwangYan}:
\begin{equation}
i {h\over 2} {\bar\Psi}_L(3,2) \times \Psi_R(3,1) \cdot
{\tilde \Phi}(3,2) + h.c.,
\end{equation}
where $\Psi(3,i)$ are the neutrino triplet just mentioned above (with
the first label for $SU_f(3)$ and the second for $SU_L(2)$). The
cross (curl) product is somewhat new \cite{Family}, referring to
the singlet combination of three triplets in $SU(3)$. The Higgs field
${\tilde \Phi}(3,2)$ is new in this effort, because it carries
some nontrivial $SU_L(2)$ charge.

Note that, for charged leptons, the Standard-Model choice is ${\bar \Psi}(\bar 3,2)
\Psi_R^C(3,1) \Phi(1,2) +c.c.$, which gives three leptons an equal mass. But, in
view of that if $(\phi_1,\phi_2)$ is an $SU(2)$ doublet then $(\phi_2^\dagger,
-\phi_1^\dagger)$ is another doublet, we could form ${\tilde\Phi}(3,2)$
from the doublet-triplet $\Phi(3,2)$. The notations in
$\Phi(1,2)$, $\Phi(3,2)$, and $\Phi(3,1)$ should be consistent
and thus the ${\tilde \Phi}(3,2)$, used in the above equation,
should have the tilde operation, for the consistency in notations.

So, we have \cite{Hwang417}
\begin{equation}
i {h^C\over 2} {\bar\Psi}_L(3,2) \times \Psi_R^C(3,1) \cdot
\Phi(3,2) + h.c.,
\end{equation}
which gives rise to the imaginary off-diagonal (hermitian) elements
in the $3\times 3$ mass matrix, so removing the equal masses of the
charged leptons. Note that the couplings $h$, $h^C$, and $\kappa$
all are dimensionless.

The expressions for neutrino oscillations and the off-diagonal mass term
are in $i\epsilon_{abc}$, or curl-dot, product - it is allowed for
$SU(3)$. Note that such coupling has nothing to do with the
kinetic-energy term of the particle, though the coupling $h$
(and $h^c$) might be related to the gauge coupling $\kappa$.

It is essential to note that all the couplings introduced
above all are dimensionless in the 4-dimensional Minkowski
space-time. This is a basic characteristic of the lepton
world in the Standard Model.

\medskip

\subsection{The Dimensionless Quark World}

We now turn our attention to the quark world, which our special
gauge-group Minkowski space-time supports. Thus, we have, for
the up-type right-handed quarks $u_R$, $c_R$, and $t_R$,
\begin{equation}
D_\mu = \partial_\mu - i g_c {\lambda^a\over 2} G_\mu^a -
i {2\over 3} g'B_\mu,
\end{equation}
and, for the rotated down-type right-handed quarks $d'_R$, $s'_R$,
and $b'_R$,
\begin{equation}
D_\mu = \partial_\mu - i g_c {\lambda^a\over 2} G_\mu^a -
i (-{1\over 3}) g' B_\mu.
\end{equation}

On the other hand, we have, for the $SU_L(2)$ quark doublets,
\begin{equation}
D_\mu = \partial_\mu - i g_c {\lambda^a\over 2} G_\mu^a - i g
{\vec \tau\over 2}\cdot \vec A_\mu - i {1\over 6} g'B_\mu.
\end{equation}

There are the standard way to generate mass for the various quarks.
For these quarks, we use the "old-fashion" way as in the Standard
Model, since quarks do not couple to the family Higgs fields. We
have, for the generation of the various quark masses,
\begin{eqnarray}
& G_1 {\bar U}_L(1,2) u_R {\tilde \Phi}(1,2) + G'_1 {\bar U}_L(1,2) d'_R
\Phi(1,2) + h.c. +\nonumber\\
& G_2 {\bar C}_L(1,2) c_R {\tilde \Phi}(1,2) + G'_2 {\bar C}_L(1,2) s'_R
\Phi(1,2) + h.c. +\nonumber\\
& G_3 {\bar T}_L(1,2) t_R {\tilde \Phi}(1,2) + G'_3 {\bar T}_L(1,2) b'_R
\Phi(1,2) + h.c.,
\end{eqnarray}
with the tilde's defined as before.

Again, all the couplings in the quark world are
dimensionless in the 4-dimensional Minkowski space-time.
Surprisingly, the natural scale for the quark world is
of fermi scales, which is five orders smaller that the
natural scale of the lepton world, of atomic scales.

It might be essential to realize that the dimensionless
couplings $g_c$, $g$, $g'$ (hence $\alpha$), and $\kappa$
(in the strengths of the fundamental interactions)
and the dimensionless mass parameters $h$, $h^C$,
$G_{1,2,3}$, $G'_{1,2,3}$ (in describing the masses
of point-like particles) have a complete equal status
in the philosophy of concepts.

The overall background, i.e., the quantum 4-dimensional
Minkowski space-time with the force-fields gauge-group
structure $SU_c(3) \times SU_L(2) \times U(1) \times
SU_f(3)$ built-in from the very beginning, supports the
"dimensionless" lepton world and it also supports the
"dimensionless" quark world. It seems that there might
be many elegant stories associated with the Standard
Model \cite{Hwang417}.

When we see the Standard Model \cite{Hwang417} from the
World of the high-enough temperature (such as from the
early Universe), it is a completely dimensionless
theory in the quantum 4-dimensional Minkowski
space-time. It is quite remarkable, indeed !!

\medskip

\subsection{The Almost-Dimensionless Overall Background}

We are in fact living in the quantum 4-dimensional
Minkowski space-time with the force-fields
gauge-group structure $SU_c(3) \times SU_L(2) \times
U(1) \times SU_f(3)$ built-in from the very beginning
\cite{Hwang417}. This is the "overall background" of
our World, or of our Universe. That is why, in our
Universe, we have $3^\circ\,K$ Cosmic Microwave
Background (CMB) as well as the clustered Cosmic
Background (CB) $\nu$'s and they were there since the
early Universe.

We may talk about "The Origin of Mass" \cite{Origin}. It stresses that,
before the spontaneous symmetry breaking (SSB), the Standard Model
does not contain any parameter that is pertaining to "mass", but, after the SSB,
all particles in the Standard Model acquire the mass terms as it should - a way
to explain "the origin of mass". Without the "ignition" term, the
Standard Model is completely dimensionless in the 4-dimensional
Minkowski space-time, including the lepton world, the quark world,
and the "overall background". In other words, when the temperature
of our Universe is high enough, the Standard Model is completely
dimensionless.

A complex scalar field in our space-time has the dimensionless coupling:
\begin{equation}
V(x) =  \lambda (\phi^\dagger(x) \phi(x))^2.
\end{equation}
The space-time integral of $L=T-V$ gives the action. In our 4-dimensional
Minkowski space-time, we find that $\lambda={1\over 8}$ numerically. This
number should come out topologically (after the normalizations of
the various fields in a given space \cite{Books}), although, at this
point, we don't know why this is the case.

If there are more than a complex scalar field, we should have
\begin{equation}
V(x) = \lambda \{(\phi_a^\dagger\phi_a)^2 + (\phi_b^\dagger
\phi_b)^2+ ....\}.
\end{equation}
There should be only one $\lambda$.

For the two related complex fields, we propose to write
\begin{equation}
V(x) = \lambda \{ (\phi_a^\dagger\phi_a + \phi_b^\dagger
\phi_b)^2 - 4 (\phi^\dagger_a \phi_b) \cdot (\phi^\dagger_b \phi_a)\},
\end{equation}
to signify the mutual attraction on top of the universal
repulsive interactions. Thus, a complex scalar field by itself
is self-repulsive and cannot exist; but with two complex scalar
fields their mutual interactions change the story.

Now we return to the Standard Model \cite{Hwang417}.
We have the Standard-Model Higgs $\Phi(1,2)$, the purely family Higgs
$\Phi(3,1)$, and the mixed family Higgs $\Phi(3,2)$, with the first label
for $SU_f(3)$ and the second for $SU_L(2)$. We need another triplet $\Phi(3,1)$
since all eight family gauge bosons are massive \cite{Family}.

It is clear that $\Phi(1,2)$ would interact with $\Phi(3,2)$ while
$\Phi(3,1)$ would also interact with $\Phi(3,2)$. These interactions
should be attractive to explain why they are showing up
together.

We could use the so-called "U-gauge" (unitary gauge). In
the U-gauge, every particle is a real particle (not a
ghost). We find it to be useful in the analysis of the
situation with the spontaneous symmetry breaking (SSB).
For the overall background, we have, in the U-gauge, $W^\pm$, $Z^0$,
and eight massive family gauge bosons, one Standard-Model Higgs and
four neutral family Higgs (three mixed plus one pure).

Thus, we choose to have, in the U-gauge, as in
\cite{Origin},
\begin{equation}
\Phi(1,2)= (0,{1\over \sqrt 2} (v+\eta)),\,\, \Phi^0(3,2) = {1\over \sqrt 2} (u_1+\eta'_1, u_2+
\eta'_2, u_3+\eta'_3 ),\,\, \Phi(3,1) = {1\over \sqrt 2}(w+\eta',0,0),
\end{equation}
all in columns. The five components of the complex triplet $\Phi(3,1)$ get
absorbed by the $SU_f(3)$ family gauge bosons and the neutral part of
$\Phi(3,2)$ has three real parts left - together making all eight family
gauge bosons massive.

To understand the origin of mass \cite{Origin}, we find that the
ignition term would better be in the purely family sector, i.e.,
the $\mu_2^2$ term. When $\mu_2^2 = 0$, the $\Phi(3,2)$ is equally
partitioned between $\Phi(1,2)$ and $\Phi(3,1)$.

It is easy to see that only one SSB-driving term is enough for
all the three Higgs fields -- there may be several SSB's for
the neutral fields - in our case, it works for all of them. SSB for
one Higgs but is driven by other Higgs - a unique feature for the
complex scalar fields. Or, we have \cite{Origin}

\begin{eqnarray}
V_{Higgs} =& \mu^2_2 \Phi^\dagger(3,1) \Phi(3,1) + \lambda
(\Phi^\dagger(1,2) \Phi(1,2)+ cos\theta_P\Phi^\dagger(3,2)\Phi(3,2))^2\nonumber\\
    &  + \lambda(-4 cos\theta_P)
(\Phi^\dagger(\bar 3,2)\Phi(1,2))(\Phi^\dagger(1,2)\Phi(3,2))
  \nonumber\\
  &+\lambda
(\Phi^\dagger(3,1) \Phi(3,1)+ sin\theta_P \Phi^\dagger(3,2)\Phi(3,2))^2
\nonumber\\ & + \lambda(-4 sin\theta_P)
(\Phi^\dagger(\bar 3,2)\Phi(3,1))(\Phi^\dagger(3,1)\Phi(3,2)).
\end{eqnarray}
These are two prefect squares minus the other extremes, to guarantee
the positive definiteness, when the minus $\mu^2_2$ was left out.

From the expressions of $u_iu_i$ and $v^2$, we obtain
\begin{equation}
v^2 (3 cos^2\theta_P-1) = sin\theta_P cos\theta_P w^2.
\end{equation}
And the SSB-driven $\eta'$ yields
\begin{equation}
w^2 (1-2 sin^2\theta_P) = - {\mu_2^2\over \lambda} +
(sin2\theta_P - tan\theta_P) v^2.
\end{equation}
These two equations show that it is necessary to have the driving
term, since $\mu^2_2=0$ implies that everything is zero. Also,
$\theta=45^\circ$ is the (lower) limit.

The mass squared of the SM Higgs $\eta$ is $2\lambda cos\theta_P u_i u_i$,
as known to be $(125\,\, GeV)^2$. The famous $v^2$ is the number
divided by $2\lambda$, or $(125\,\,GeV)^2/(2\lambda)$. Using PDG's for
$e$, $sin^2\theta_W$, and the $W$-mass \cite{PDG}, we find
$v^2=255\,\, GeV$. So, $\lambda={1\over 8}$, a simple model indeed.

The ratio of the VEV to its Higgs mass is determined
by $2\lambda$, whether the channel is not ignited or not. We might
choose the channel of $\eta'$ (the purely family Higgs) or that of
$\eta$ (the SM Higgs) as the ignition channel, but three Higgs channels
have different labels. The three Lorentz-invariant scalar fields have
different internal structures - an amusing question for further
investigation.

The mass squared of $\eta'$ is $-2(\mu_2^2-sin\theta_P u_1^2 +
sin\theta_P (u_2^2+u_3^2))$. The  other condensates are $u_1^2= cos\theta_P v^2
+ sin\theta_P w^2$ and $u_{2,3}^2 = cos\theta_P v^2 - sin\theta_P w^2$ while
the mass squared of $\eta'_1$ is $2\lambda u_1^2$, those of $\eta'_{2,3}$ be
$2\lambda u_{2,3}^2$. The mixings among $\eta'_i$ themselves are neglected
in this paper.

There is no SSB for the charged Higgs $\Phi^+(3,2)$. The mass
squared of $\phi_1$ is $\lambda(cos\theta_P v^2 - sin\theta_P w^2) + {\lambda\over 2}
u_i u_i$ while $\phi_{2,3}$ be $\lambda(cos\theta_P v^2 + sin\theta_P w^2)
+ {\lambda\over 2} u_i u_i$. (Note that a factor of ${1\over 2}$ appears
in the kinetic and mass terms when we simplify from the complex case to
that of the real field; see Ch. 13 of the Wu-Hwang book \cite{Books}.)

A further look of these equations tells that $3cos^2\theta_P - 1 > 0$ and
$2sin^2\theta_P -1 > 0$. A narrow range of $\theta_P$ is allowed (greater
than $45^\circ$ while less than $57.4^\circ$, which is determined by
the group structure). For illustration, let us choose
$cos \theta_P = 0.6$ and work out the numbers as follows:
(Note that $\lambda={1\over 8}$ is used.)
\begin{eqnarray}
& 6 w^2 = v^2, \quad -\mu^2_2/\lambda = 0.32 v^2;\nonumber\\
\eta: & 2\lambda cos\theta_0 u_i u_i =(125\, GeV)^2, \quad v^2 = (250\,GeV)^2;
\nonumber\\
\eta': & mass^2 = (51.03\,GeV)^2, \quad w^2=v^2/6; \nonumber\\
\eta'_1: & mass^2= (107\,GeV)^2, \quad u_1^2=0.7333 v^2; \nonumber\\
\eta'_{2,3}: & mass^2 = (85.4\,GeV)^2, \quad u_{2,3} = 0.4667 v^2; \nonumber\\
\phi_1:& mass = 100.8\, GeV; \qquad \phi_{2,3}: mass = 110.6\,GeV.
\end{eqnarray}
All numbers appear to be reasonable. In the above, $cos\theta_P$ is the
only free parameter until one of the family Higgs particles $\eta'_{1,2,3}$
and $\eta'$ is found experimentally. Since the new objects need to be
accessed in the lepton world, it would be a challenge for our experimental
colleagues.

{\it As a footnote, our Standard Model predicts that the mass of the SM
Higgs $\eta$ is a half of the vacuum expectation value $v$ - a prediction
in the origin of mass \cite{Origin}.}

As for the range of validity, ${1\over 3} \le cos^2\theta_P \le {1\over 2}$.
The first limit refers to $w^2=0$ while the second for $\mu_2^2 = 0$.

We may fix up the various couplings, using our common senses. The
cross-dot products would be similar to $\kappa$, the basic coupling of
the family gauge bosons. The electroweak coupling $g$ is
$0.6300$ while the strong QCD coupling $g_s=3.545$; my first guess
for $\kappa$ would be about $0.1$. The masses of the family gauge
bosons would be estimated by using ${1\over 2}\kappa \cdot w$, so
slightly less than $10\,GeV$. (In the numerical example with $cos
\theta_P=0.6$, we have $6 w^2= v^2$ or $w=102\,\,GeV$. This gives
$m=5\,\,GeV$ as the estimate.) So, the range of the family forces,
existing in the lepton world, would be $0.02\,\, fermi$.

In \cite{Origin}, the term that ignites the SSB is chosen to be with
$\eta'$, the purely family Higgs. This in turn ignites EW SSB
and others. It explains the origin of all the masses, in terms
of the spontaneous symmetry breaking (SSB). SSB in $\Phi(3,2)$
is driven by $\Phi(3,1)$, while SSB in $\Phi(1,2)$ from the
driven SSB by $\Phi(3,2)$, as well. The different, but related,
scalar fields can accomplish so much, to our surprise.

We note that, at the Lagrangian level, the $SU_c(3) \times SU_L(2)
\times U(1) \times SU_f(3)$ gauge symmetry is protected but the symmetry
is violated via spontaneous symmetry breaking (via the Higgs mechanisms).

We iterate that the mathematics of the three neutral Higgs, $\Phi(1,2)$
(Standard-Model Higgs), $\Phi(3,1)$ (purely family Higgs), and
$\Phi^0(3,2)$ (mixed family Higgs), subject to the renormalizabilty
(up to the fourth power), turns to be rather rich. In our earlier work
regarding the "colored Higgs mechanism" \cite{HwangWYP}, we show how
the eight gauge bosons in the $SU(3)$ gauge theory become massive using two
complex scalar triplet fields (with the resultant four real Higgs fields),
with a lot of choices. We suspect that, even within QCD, there might be
some elegant choice of "colored" Higgs, or there must be a good reason
for massless gluons.

\medskip

\subsection{The Standard Model as a dimensionless theory}

In the $\hbar = c =1$ unit system, the Standard Model
\cite{Hwang417}, apart from the "ignition" term, is a
collection of a few pure numbers; that all couplings
are dimensionless. That is, in each interaction, the
overall dimension of all field variables just cancels
out the overall dimension of the 4-dimensional
Minkowski space-time.

This is a remarkable property of the Standard Model.
We should follow this line to examine the problem of
infinities (i.e. ultraviolet divergences). So, the
Standard Model is basically a dimensionless theory;
in which the term "mass" still does not assume the
meaning \cite{Origin}.

To re-iterate what what we have for the Standard
Model: The lepton world is
dimensionless in the 4-dimensional Minkowski
space-time. The quark world is also dimensionless
in the 4-dimensional Minkowski space-time. Except the
SSB "ignition" term, the overall background is also
dimensionless in the 4-dimensional Minkowski space-time.
"Dimensionless in the 4-dimensional Minkowski space-time"
might mean that it is determined {\it globally} by the
quantum 4-dimensional Minkowski space-time.

In other words, a set of pure numbers in the 4-dimensional
Minkowski space-time, $\{ g, g', \kappa, h, h^c, ... \}$,
characterizes the lepton world, there is another set of
pure numbers, $\{g_c, g, g', G_1, ..., G'_1, ...\}$,
defines the quark world, and $\{ \lambda, \theta_P\}$ and
$\mu_2^2$ characterize the overall background. When the
temperature is high enough (in comparison with
$-\mu_2^2$), it is a completely pure-number game
in the quantum 4-dimensional Minkowski space-time.

This may lead to a complete ball game for the
overall treatment of ultraviolet divergences -
the headache problem of infinities
\cite{Books, fine-tune}.

Thus, the Standard Model is a dimensionless
theory \cite{Hwang417}, if we forget about the
"ignition" term or, equivalently, if we go to the
temperature high enough. Just looking at those
integrals for $m=0$, we have some
yet-to-be-defined integrals to investigate.
Anyway, the Standard Model, as a dimensionless theory,
should be used, by mathematicians or theoretical
physicists, to investigate the problem of
ultraviolet divergences. Upon the SSB or the
generalized Higgs mechanism, the problem becomes
so involved and so complicated. Thus, we should
deal directly with the Standard Model in the
dimensionless phase; there the mass still does
not assume its meaning \cite{Origin}.

\medskip

\subsection{Phase Transitions in View of the Standard Model}

If we look at the Standard Model before the spontaneous
symmetry breaking (SSB), it is a completely dimensionless
theory \cite{Hwang417} - i.e., a complete collection of
pure numbers in the quantum 4-dimensional Minkowski
space-time. Before SSB, "mass" does not assume its
meaning since there are no mass terms.

The critical temperature seems to play a similar 
role as the spontaneous symmetry breaking (SSB). 
In the early Universe, the temperature lowered 
as the time passed - at the critical temperature 
$T_c$, a phase transition of certain sort takes 
place. But SSB happens because of some negative 
energy that is "igniting" from the hand of the 
God. This sounds like some pure imaginary 
temperature working for us; in reality, we 
can not tune our environment in that way.

When we are talking about the Standard Model above
the critical temperature $T_c$, we are dealing,
{\it effectively} with a dimensionless theory in 
the quantum 4-dimensional Minkowski space-time. 
Then, all the masses become vary small and {\it 
effectively} could be neglected. This is still 
very different from the imaginary temperature 
(such as SSB). 

At the time approximately $10^{-11}\,sec$ in
the early Universe, our Universe underwent the
so-called "mass phase transition". Just after
the mass phase transition, our Universe remained
as the "quark soup" till a little after
$10^{-5}\,sec$ - then, it experienced the
cosmological QCD phase transition. From there
on, the clustering processes set in, first in
abundance in cosmic microwave background
(CMB), cosmic background (CB) $\nu$'s,
electrons, and the light elements (such as
$p$, $d$, etc.), through the dark age, and
finally resulting in formation of stars
and galaxies.

When the density is so high such as being
compatible even in the Schwarzschild metric,
the term in "mass"should have to be
re-interpreted otherwise. In our
real Universe, each visual ordinary-matter object,
such as the Earth, the Venus, the Sun, etc.,
should have the $25\%$ dark matter, which is
recently identified as a neutrino halo
\cite{HwangP1}. Neutrinos are fermions, possessing
the Fermi-Dirac sphere \cite{Hwang9}. The surface
Fermi energy would be $600\,GeV$ for the neutrino
halo of the Earth \cite{Hwang8} - the large Fermi
energy is due to the tiny mass(es) of the
neutrinos. So, the system is jacking up via Pauli's
exclusion principle (i.e., the quantum principle)
while the tiny size related to the Schwarzschild
metric (i.e., for black holes) becomes irrelevant.

On the other hand, when we are talking about
infinities (ultraviolet divergences), if it is
coming from a theory with the mass terms
($m\ne 0$) or it is from a completely massless
theory, the distinction may be rather
fundamental. Thus, in the early Universe, there
is the period with a completely massless theory
(above the critical temperature $T_c$); then,
eventually, with a theory with the meaningful
mass terms. The treatments of the various
ultraviolet divergences in the two cases are
completely different.

We wish to make a fundamental note that the phase
transitions of the Standard Model are "irreversible"
in the sense of thermodynamics. The sign of the
"ignition" term is negative in the Standard Model
\cite{Hwang417} and, so, it can be
achieved only through the hand of the God.

{\it So, the phase transition at high temperature 
of the Standard Model only has the imaginary image
of the mass "phase transition", or SSB, of the 
Standard Model.}

\medskip

\section{Our Universe}

{\it Our Universe is defied as the world described
by the Standard Model \cite{Hwang417}.} So, our
Universe is the quantum 4-dimensional Minkowski
space-time with the force-fields gauge-group
structure $SU_c(3) \times SU_L(2) \times U(1)
\times SU_f(3)$ built-in from the very beginning.
The Universe accepts the lepton world, of atomic
sizes. The Universe also accepts the quark
world, of nuclear sizes.

Our Universe has, as its contents, the $5\%$ visual
ordinary matter, the $25\%$ dark matter, and the
$70\%$ dark energy. According to the Standard
Model, electrons, quarks, and photons are
visual ordinary-matter particles while
neutrinos (of three flavors, and all antineutrinos)
are the {\it only long-lived} invisible particles.
The $70\%$ dark energy is probably uniformly
distributed in our Universe.

In terms of the smallest units of matter, a
star of five solar mass would be an aggregate
of $10^{60}$ smallest units of matter (in terms of
the number of electrons) - a gigantic number!
The approximate Newton's gravitational law
is used to describe the gravity between two
aggregates of such gigantic numbers of the
smallest units of matter.

This gigantic number of $10^{60}$ (in the
smallest units of matter) does tell us a lot
of things. For example, the theory of
gravity is far from the Standard Model.
The existence of the smallest units of
matter carries the implication that
mutual unification is among $SU_c(3)
\times SU_L(2) \times U(1) \times SU_f(3)$
and that it should not extend to cover
the theory of gravity.

Neutrinos (hereafter used to represent "neutrinos
of all three flavors, and all antineutrinos"),
together with photons, were produced in abundance
in the early Universe, forming the Cosmic Microwave
Background (CMB) and the Cosmic Background (CB)
$\nu$'s.

CB$\nu$'s, owing to the neutrino mass, tend to
cluster, forming the so-called "neutrino halos",
each for a visual ordinary-matter object such as
the Earth, or the Venus, or the Sun (the star).
Because the mass of the neutrino is tiny, the
neutrino halo should not exist by itself.

If we examine the Standard Model closely, we
could safely conclude that neutrinos would be
the {\it only} long-lived dark-matter particles
\cite{HwangP1} and that the $25\%$ dark matter
should be the clustered neutrino halos
\cite{Hwang9, Hwang8}.

So, the early stage of our Universe should have
the very early Universe, which was not yet
clusterized, and the clustered early Universe,
which already exhibited the lumps and clusters.
The (dark-matter) neutrino halos and the lumps
of the visual ordinary-matter objects are the
major products, during the clustered early
Universe.

If we look at Newton's universal gravitational
law,
\begin{equation}
m a = force = G'_N {m M\over r^2},
\end{equation}
the $m$ could be applied to some visible object
or to some invisible object (though not seen by
us), and the $M$ should also contain the visual
part in ordinary matter and the invisible part
in dark matter (neutrino halo). Since a neutrino
halo cannot form the center of weight (because of
the tiny masses of neutrinos) and it has to follow
a visual ordinary-matter object (as the center of
weight). This is the consequence of Newton's
universal gravitational law.

The existing prevailing view regarding our Universe
is that it originates from the joint making of
Einstein's general relativity and the cosmological
principle. Based upon the cosmological principle
which state that our universe is homogeneous
and isotropic, we use the Robertson-Walker
metric to describe our Universe \cite{Turner}.
\begin{equation}
ds^2=dt^2 -R^2(t)\{ {dr^2\over 1-kr^2} +r^2 d\theta^2
+r^2 sin^2\theta d\phi^2\}.
\end{equation}
Here the parameter $k$ describes the spatial curvature with
$k=+1$, $-1$, and $0$ referring to an open, closed, and flat
universe, respectively. The scale factor $R(t)$ describes the size
of the universe at time $t$.

Assuming that the very early Universe can
be described by a perfect fluid, i.e., a
fluid with the energy-momentum tensor
$T^\mu \ _\nu = \, diag\, (\rho,\ ,\ -p,\ -p,\, -p)$ where $\rho$
is the energy density and $p$ the pressure (both as functions of
$t$), then the Einstein
equation, $G^\mu\ _\nu = 8\pi G_N T^\mu\ _\nu + \Lambda g^\mu\
_\nu$, gives rise to only two independent equations, i.e., from
$(\mu,\ \nu) = (0,\ 0)$ and $(i,\ i)$ components,
The two equations can be combined to yield
\begin{equation}
{\ddot R\over R} = -{4\pi G_N\over 3} (\rho +3 p) + {\Lambda \over 3}.
\end{equation}
This last equation shows either that there is a
positive cosmological constant or that $\rho + 3p$
must be somehow negative, if the major conclusion
of the Supernovae Cosmology Project is correct \cite{SNIa}, i.e.
the expansion of our universe still accelerating (${\ddot R/ R}
>0$).

Alternately, assuming a simple equation of state,
$p= w \rho$, we obtain, from the two independent
equations,

\begin{equation}
2 {\ddot R\over R}+ (1+ 3w)({\dot R^2\over R^2} + {k\over R^2})- (1+w)\Lambda=0,
\end{equation}
which can easily be solved for a constant $w$.

For example, let us assume for $t \ge T_0$ ($T_0$
a few times the age of our Universe) that it is
dominated by the matter with a constant $w_0$.
We could write
\begin{equation}
R=e^{-at}+ ce^{at},
\end{equation}
with the proportional constant fixed up if necessary. We obtain
\begin{equation}
a^2 = \Lambda, \qquad c = {k\over {4a^2}}.
\end{equation}
Note that we try to fix the constants
$a^2$ and $c$ in the asymptotic solution.

It is interesting to note, in this
asymptotic solution ($t\to +\infty$), that
(1) $w$ is dropped out completely,
(2) $\Lambda$ is better to be nonnegative
(and very small), (3) to describe an
expanding Universe, $a$ should be negative,
and (4) it is better to be the case that
$c k \ge 0$. Asymptotically, this would
be the solution as $t \to \infty$.

The above situation applies to $t\to \infty$
and NOT in the very early Universe.
For the very early Universe, we may set
$\Lambda$ and $k$ both to zero. Or, we have
\begin{equation}
2 {\ddot R}R + (1+3w){\dot R}^2 = 0, \qquad or,\, R\propto t^n, \quad n={2\over 3(1+w)}.
\end{equation}
The exception is at $w=-1$, for the inflation era.
To describe the Inflation Era, we use $p= -\rho$ so that
\begin{equation}
\ddot R - {\dot R^2\over R}=0,
\end{equation}
which has an exponentially growing, or decaying, solution $R
\propto e^{\pm \alpha t}$, compatible with the so-called
"inflation" or "big inflation". In fact, considering the simplest
case of a real scalar field $\phi(t)$, we have
\begin{equation}
\rho = {1\over 2} {\dot \phi}^2 + V(\phi), \qquad p ={1\over 2} {\dot \phi}^2 -V(\phi),
\end{equation}
so that, when the "kinetic" term ${1\over 2}{\dot \phi}^2$ is
negligible, we have an equation of state, $p \sim -\rho$. In
addition to its possible role as the "inflaton" responsible for
inflation. Such field has also been invoked to explain the
accelerating expansion of the present universe, as dubbed as
"quintessence" or "complex quintessence" \cite{Quint}.

Let's look at the standard textbook argument leading to the
radiation-dominated universe and the matter-dominated universe:

For the {\it Radiation-Dominated Universe}, we have $p=\rho/3$.
For simplicity, we assume that the curvature is zero ($k=0$) and
that the cosmological constant is negligible ($\Lambda =0$). In
this case, we find from Eq. (23)
\begin{equation}
R \propto t^{1\over 2}.
\end{equation}

Another simple consequence of the homogeneous model is to derive
the continuity equation:
\begin{equation}
d (\rho R^3) +p d(R^3)=0.
\end{equation}
Accordingly, we have $\rho\propto R^{-4}$ for a
radiation-dominated universe ($p = \rho/3$) while $\rho \propto
R^{-3}$ for a matter-dominated universe ($p << \rho$).  The
present universe is believed to have a matter content of about
5\%, or of the density of about $5 \times 10^{-31} g/cm^3$, much
bigger than its radiation content $5 \times 10^{-35} g/cm^3$, as
estimated from the $3^\circ$ black-body radiation. However, as $t
\to 0$, we anticipate $R \to 0$, extrapolated back to a very small
universe as compared to the present one. Therefore, the universe
is necessarily dominated by the radiation during its early enough
epochs.

For the radiation-dominated early epochs of the universe with
$k=0$ and $\Lambda =0$, we could deduce,
\begin{equation}
\rho ={3\over 32 \pi G_N}t^{-2},\qquad T =\{ {3c^2\over 32 \pi G_N a}\}^{1\over 4}
t^{-{1\over 2}} \cong 10^{10}t^{-1/2} (^\circ K).
\end{equation}
These equations tell us a few important times in the early universe, such as $10^{-11} sec$
when the temperature $T$ is around $300\ GeV$ during which the electroweak (EW) phase
transition is expected to occur, or somewhere between $10^{-5} sec$ ($\cong 300\ MeV$)
and $10^{-4}sec$ ($\cong 100\ MeV$) during which quarks and gluons undergo the QCD
confinement phase transition.

For the {\it Matter-Dominated Universe}, we have $p\approx 0$,
together with the assumption that $k=0$ and $\Lambda=0$. Eq. (23)
yields
\begin{equation}
R \propto t^{2\over 3}.
\end{equation}
As mentioned earlier, the matter density $\rho_m$ scales like
$R^{-3}$, or $\rho_m \propto t^{-2}$, the latter similar in the
radiation-dominated case.

When $t=10^9 sec$, we have $\rho_\gamma=6.4\times 10^{-18}gm/cm^3$
and $\rho_m=3.2\times 10^{-18}gm/cm^3$, which are close to each
other and it is almost near the end of the radiation-dominated
universe. The present age of the Universe is 13.7 billion years -
for a large part of it, it is matter-dominated although now we
have plenty of dark energy (65\% $\sim$ 70\%).

However, it is generally believed that our present universe is
already dominated by the dark energy (the simplest form being of
the cosmological constant; about 70\%) and the dark matter (about
25\%). The question is when this was so - when the dark part
became dominant.

In other words, the proper language should include (1) the radiation
density, (2) the matter density, (3) the dark-matter density, and (4)
the dark-energy density - and it determines the evolution of the Universe.
Apparently, the entry of dark matter and of dark energy only makes the study
of evolution of the Universe much more interesting.

\medskip

\section{The Mathematical Domain of the Lorentz Symmetry}

There is a fundamental question regarding the mathematical
domain of the space-time coordinates $(x,y,z,ct)$ - are they
the pure numbers or the partially-commuting operators?
Of course, we could always implement the Lorentz symmetry.

As a matter of fact, we did not ask ourselves whether
the coordinates $(x, y, z, ct)$ are four pure numbers.
This would be the basic question that we're trying to ask.
In fact, the answer is that we do not know.

Usually, we go to the extreme limits when we start asking
such question. In fact, the question is always there, except
that we are used to the "normal" working situations and then
forgetting the right to ask.

What do we anticipate well above the critical temperature
$T_c$ of the mass phase transition? The Standard Model
\cite{Hwang417} describes electrons, neutrinos, quarks,
etc., so well. The Standard Model should yield the
equation of state ($EoS$), which we could use for the
problem.

The other important question is what we might have to do
by going to the extremely high matter densities, while
keeping the Lorentz symmetry. In what
follows, we try to figure out if we are hitting the
so-called {\bf "super-quantum"} regime. Let us try to
explain the underlying idea, while remembering that
the Standard Model and the Lorentz symmetry works
extremely well. (We may call it "super-quantum" or
"sub-quantum"; we prefer to use "super-quantum" in what
follows.)

Suppose that the formulae in the last section are correct; then at time
$t = 10^{-15} sec$ we could determine $T$, $\rho_m$, and $\rho_\gamma$
as follows:

\begin{equation}
T= 32\, TeV: \quad \rho_m=3.2 \times 10^{18} gm/cm^3, \quad \rho_\gamma =
6.4 \times 10^{30} gm/cm^3.
\end{equation}
To get the feeling of these numbers, suppose that we stack the entire Solar
mass into a small sphere that it becomes a black hole: $\rho_m= 1.843 \times 10^{16}
gm/cm^3$ at $R_s= 2.953 km $. The nuclear matter density $\rho_N$ is $2\times
10^{14} gm/cm^3$.

The "mass" no longer has its meaning; $\rho_m$
has to be re-interpreted as some form of "clustered
energy" - visible rather than invisible. But
$\rho_\gamma$, the energy carried by the bulk of
photons, does not suffer from the same ambiguity.

So, at time $10^{-15} sec$, the "matter" density
$3.2\times 10^{18} gm/cm^3$ is
already many orders of magnitude denser (or higher) than
that below the critical temperature $T_c$ of the mass phase
transition (at about $10^{-11}\, sec$). But this is what is
predicted by our standard theory (i.e. the
Friedman-Robertson-Walker metric and the equation
of state for ordinary matter). Of course, the radiation
density (the boson's density) is even more ridiculously
higher.

At this point, we have some ideas about $\rho_{DM}$
of our Universe;
presumably it all depends on how the majority of
dark matter gets manufactured - neutrino halos as
the $25\%$ dark matter \cite{Hwang9, Hwang8}. So,
neutrino halos scale like the visual ordinary-matter
objects, such as the Earth, the Sun, and stars.

On the dark energy of our Universe, it would
be tiny at time $t=10^{-15}\, sec$, if it is
indeed described by the cosmological constant,
and alike.

At this juncture we could remind ourselves the so-called "quantum regime",
where the momenta $\vec p$ do not commute with the coordinates $\vec x$.
The quantum regime is relevant when we deal with ordinary atomic or subatomic
scale (or, distances). When the Universe was $10^{-15}$ second old (or,
earlier), we are dealing with sizes much smaller than
the size of each individual hadron - it might be safe to introduce
the "super-quantum regime": At these scales, the basic variables
are still the coordinates $(\vec x,\, t)$ but we could generalize
them by making them non-commuting, or operators, by keeping
intact the Lorentz symmetry. (What else?)

The presence of the quantum regime at the atomic or subatomic scale
suggests the "existence" of the "super-quantum" regime, perhaps
$10^{-20}\,cm$ away. Again, the concept of "point-like" in physics
could be different from the ideological mathematical "point-like".
In fact, the thoughts are similar to those suggested from a lot of
physicists, famous or not. It is the very fundamental difference
between the physicists and the mathematicians, such as the meaning
of "point-like-ness" and others.

Why do we have to worry about the super-quantum regime well
ahead of the Planck era? To say that from the Planck length
$1.616\times 10^{-33} cm$ to $10^{-18} cm$ ($10^{-5}fm$),
or from the Planck time $10^{-43}sec$ to
$10^{-15} sec$, there is nothing (desert) -
nobody would believe in it. In other words, we should
worry about the "super-quantum" era if we analyze
the "size" and the matter together and the
"contradiction" (or the paradox) that we have.

The "size" of an object, such as an electron or a hadron, is not
a well-defined concept. LEP at CERN allowed us to accelerate the electrons
to $100 \,GeV$, or let us probe $10^{-6} fermi$ without the need to take
into account the "size" of the electron. But how small could we go in
this direction? For hadrons, the "size" is dictated by the strong
interactions, or QCD. Analogously, maybe the "size" of an electron
is dictated by the electroweak theory, in the $TeV$ regime. Maybe
the "super-quantum" regime could be set in after that, or orders after
that, but definitely well before the Planck scale.

It is well-known from the textbooks on quantum statistical mechanics that
the scale or distance for the spacing of particles should be larger than
the intrinsic distance of the individual particle (or the de Broglie wavelength
of the molecule) - clearly, we eventually would go to the regime that this very
basic fact would be contradicted in defining the quantum statistical
mechanics. In other words, we have no basis to "derive" the
equation of state (EoS), to be used together with the Einstein equation
(on general relativity, something alike). In our opinion, we should try
to face this difficult situation instead of entertaining ourselves that
we already have solved a complete set of equations for cosmology.

What would be the algebra, or the graded Lie algebra, among $(x,\,y, \,z )$
or among $(x,\,y,\,z,\, ict)$? Let's try to argue in the following way. In going
from the "classical" regime to the "quantum" regime, we make the quantities
$({\vec x},\,{\vec p})$ noncommutative and thus obtain the operator algebra.
Now in the "quantum" regime, the basic variables are $(x,\,y,\,z,\,ct)$ and the
most general, and perhaps the only, way in the "super-quantum" regime is trying
to introduce some noncommutative algebra among these variables $(x,\,y,\,z,\,ct)$
while keeping the Lorentz symmetry.

As mentioned earlier, the "matter" density is predicted to be
$3.2\times 10^{18} gm/cm^3$ when the Universe was $10^{-15} sec$. This means
that a volume of one hadron size (judging from the nuclear matter density) would
accommodate 16,000 hadrons or matter particles. (Some people wouldn't worry
about the situation until the Planck time or the Planck distance is reached.)
Our thinking is as follows: As we go to smaller and smaller such that a volume
of one hadron size has to accommodate thousands or more hadrons, the size by
itself does not assume the meaning and instead the size and the matter may
jointly have the meaning. The size without the matter, or the matter without
the size (defined through its space-time), maybe lacks the (physics) meaning.

In the simple case, we need to take the coordinates $(x,\,y,\,z,\,ct)$ as from
the operators and to the first approximation to implement Lorentz symmetry
in some way. Since Lorentz invariance is known to be true to our
experimental accuracy, we are lucky to have this important guideline. The next
question is to define the "operations" among ${\hat x}$, ${\hat y}$,
${\hat z}$, and $c{\hat t}$ such that $d^2s$ and others may have
their meanings.

We could put more thoughts along this line. Let $(x,\,y,\,z,\,ct)$ be the Cartesian
variables, which we need to label the coordinates (operators).
The coordinates $({\hat x},\,{\hat y},\,{\hat z},\,c{\hat t})$ may be regarded
as mappings from $R^4$ to $C^4$ (or something else). The next thing is to
implement Lorentz symmetry - to define the ten operators, i.e. the Hamiltonian $H$,
the three momentum $\vec P$, the three angular momentum $\vec J$, and the three
Lorentz-boost operators $\vec K$ or $\vec M$. These ten operators should be the mappings
from the coordinates (operators themselves). Fortunately, the idea was already
suggested much early on by H. Snyder \cite{Snyder}, for a different basic
motivation.

In other words \cite{Snyder}, $\hat x$, $\hat y$, $\hat z$, and $\hat t$ are
hermitian operators for the space-time coordinates of a particular Lorentz frame;
the operators $\hat x$, $\hat y$, $\hat z$, and $\hat t$ are such that the
spectra of the operators $\hat x',\,\hat y',\,\hat z',\,\hat t'$ formed by
taking linear combinations of $\hat x,\,\hat y,\,\hat z,\,\hat t$, which
leaves invariant the quadratic form, $s^2 = x^2 + y^2 + z^2 - c^2 t^2$,
shall be the same as the spectra of $\hat x$, $\hat y$, $\hat z$, and $\hat t$.

To find operators $\hat x$, $\hat y$, $\hat z$, and $\hat t$ possessing
Lorentz invariant spectra, we may consider the homogeneous quadratic
form \cite{Snyder}:
\begin{equation}
\eta^2 = -\eta_0^2 + \eta_1^2 + \eta_2^2 + \eta_3^2 + \eta_4^2,
\end{equation}
in which the $\eta$'s are assumed to be real variables and they may be
regarded as the homogeneous projective coordinates of a real four-dimensional
space of constant curvature (a de Sitter space).

We may define \cite{Snyder}
\begin{equation}
\hat x = i a (\eta_4 \partial /\partial \eta_1 - \eta_1\partial /\partial \eta_4),
\end{equation}
and analogously for $\hat y$, $\hat z$, and $c \hat t$, where $a$ is some unit
length (defined in the super-quantum regime; it was the nature unit in the Snyder's
discrete space-time). Accordingly, we introduce three angular-momentum
operators and three Lorentz-boost operators as follows:
\begin{equation}
L_x= i \hbar (\eta_3\partial /\partial \eta_2 - \eta_2 \partial /\partial \eta_3)
\end{equation}
and analogously for $L_y$ and $L_z$.
\begin{equation}
M_x = i \hbar (\eta_0\partial/\partial \eta_1 + \eta_1 \partial/\partial \eta_0)
\end{equation}
and similarly for $M_y$ and $M_z$. Thus, we have
\begin{equation}
[\hat x,\, \hat y] = (i a^2/\hbar) L_z, \quad [c \hat t,\,\hat x] = (i a^2/\hbar) M_x,
\end{equation}
and so on (noncommutative geometry in terms of six relations).

There may exist an interesting connection for this de Sitter space. We know
that the cosmology written on the 4-dimensional hyper-surface $-\eta_0^2 +
\eta_1^2 + \eta_2^2 + \eta_3^2 + \eta_4^2 = \alpha^2$ would possess the
cosmological constant $\Lambda= 3/\alpha^2$. In the Snyder's language
the interchange $\eta_4 \to \eta$ is in fact optional. Therefore, we have
the most basic "prediction" - the cosmological constant $\Lambda$ is connected
with the physics of the "super-quantum" era. Now we have \cite{PDG}
\begin{equation}
\rho_c \Omega_\Lambda = 7.20565\times 10^{-30} gm \cdot cm^{-3} =
4.0421 \times 10^{-6} (GeV/c^2)\cdot cm^{-3},
\end{equation}
which explains the current dark energy density (74\% of the critical energy density).
Equating this to $\Lambda/(8\pi G_N)$, we obtain
\begin{equation}
\Lambda = 1.2087 \times 10^{-35} sec^{-2}.
\end{equation}
Hence, we have
\begin{equation}
\alpha^2 = 3/\Lambda = 2.4820 \times 10^{35} sec^2,\quad or \quad \alpha = 4.98
\times 10^{17} sec.
\end{equation}
It is quite amazing to be close to the age of our Universe: 13.69 Gyr ($=4.3202
\times 10^{17}\,sec $). Since what we do is to specify a four-dimensional surface
in a de Sitter space, these numbers could be taken as the consistency of these
thoughts.

Surprisingly enough, the value of de Sitter $\alpha$ as derived from the tiny
$\Lambda$ (previously coming from nowhere) is remarkably close to the age of
our Universe - this might take as the clue to resolve the problems of small
cosmological constant, etc.

Our discussion indicates that there are other ways to introduce the noncommutative
algebra \cite{Snyder,Connes} but there are reasons to try out the Snyder's option.

The concrete example \cite{Snyder} which we try to show, in light of
the Standard Model \cite{Hwang417}, is that, assuming Einstein's
relativity principle and the quantum principle, the underlying algebra
still has a big room to vary. Assuming even in addition that there
exist the smallest units of matter such as electrons, neutrinos,
quarks, etc. (a finite of them), we are not so sure if the book
of Snyder is already closed. In our own opinion, thinking of this
kind is the right way which the physics should go.

\medskip

\section{On the Cosmological QCD Phase Transition}

After presenting the perfect phase transition, the mass phase
transition of the Standard Model \cite{Hwang417}, the room left
for the QCD cosmological phase transition \cite{Hwang0} seems to
be rather limited. But it is an unsolved problem so far.

In an early paper \cite{Hwang0}, we present our thoughts about the cosmological QCD
phase transition. It happened at between
$10^{-5}sec$ and $10^{-4} sec$, when the densities appear to be "normal"
for the hadron matter. That is, we don't have to worry about "super-quantum"
physics. However, the real story about the QCD phase transition
is rather complicated but may be truly important for phase
transitions in the early Universe. The question in mind has to do with
the latent heat (energy) of the phase transition of first-order - similar
to the zero-point energy, or the Cosmological Constant, or that, when we
differentiate, it disappears; supposedly that it is no longer relevant but it is
not true.

To describe the cosmological QCD phase transition, it is unlikely that we
would wait for until we have solved QCD itself - often a formidable task.
In this case, we could use bag models as the first approximation
but maybe the phase transition in question is described as the first-order
transition while in reality a second-order one. Then, solving QCD exactly
becomes very important. Of course, we all know that the MIT bag model \cite{MIT}
or the Frieberg-Lee non-topological soliton model \cite{TDLee} invokes the bag
constant, or the "zero-point" energy, which implies the first order for phase
transition.

In what follows, we use the "bubble dynamics" in Frieberg-Lee
non-topological soliton model - to describe what is going
on in the formation and evolution of many bubbles, trying
to model the dynamics in the cosmological QCD phase transition.

At the temperature $T>T_c\sim 150 MeV$, i.e., before the phase
transition takes place, free quarks and gluons can roam anywhere.
As the Universe expands and cools, eventually passing the critical
temperature $T_c$, the bubbles nucleate here and there. These
bubbles "explode", as we call it "exploding solitons" (or "low-temperature
solitons"). When it
reaches the "supercooling" temperature, $T_s$, or something
similar, the previous bubbles become too many and in fact most of
them become touched each other - now the false vacua or "bubbles"
of different kind (where quarks and gluons can move freely) start
to collapse - or we call it "imploding solitons" (or
"high-temperature solitons"). When all these
bubbles of different kind implode completely, the phase transition
is now complete.

The "imploding" solitons with boundaries should find a way to "glue"
together, such as the formation of domain walls, vortices, etc., sometimes
with nontrivial topology.

There is some specialty regarding the cosmological QCD phase
transition. Namely, the collapse of the false vacuum does depend
on the inside quark-gluon content - e.g., if we have a three-quark
color-singlet combination inside, the collapse of the false vacuum
would stop (or stabilize) at a certain radius (we called the bag
radius, like in the MIT bag radius); of course, there are meson
configurations, glueballs, hybrids, six-quark or multi-quark
configurations, etc. The cosmological QCD phase transition does
not eliminate all the false vacua; rather, the end state of the
transition could have at least lots of baryon or meson states,
each of them has some false vacuum to stabilize the system.

How big can a bubble grow? It is with the fastest speed which the
bubble can grow is through the speed of light or close to the
speed of light. The bubble could sustain from the moment it
creates, say, $T\approx T_c$ to the moment of supercooling,
$T_s\sim 0.95 \cdot T_c $, or during the time span $t \sim 3\times
10^{-5} \times 0.05 sec $ (or $1.5 \times 10^{-7} sec$). So, the
bubble can at most grow into $c \cdot 1.5 \times 10^{-7} sec$ or
$4.5 \times 10^{3}\, cm$.

How many ("low-temperature") bubbles are there of
the entire Universe when the space were filled
up by the bubbles (when the phase transition was complete)?
The point is that two bubbles are separated by the domain wall of
certain structure (with some energy deposited in there - some
surface energy). The domain walls cannot disappear completely
- not only sometime because of the possible nontrivial topology
but that there should be some QCD dynamics to annihilate
the walls.

As a yardstick, we note that, at $t \sim 10^{-5}\,sec$ or $T\sim
300\,MeV$, we have
\begin{equation}
\rho_\gamma\sim 6.4 \times 10^{10}gm/cm^3,\qquad \rho_m\sim
3.2\times 10^3 gm/cm^3.
\end{equation}
Or, slightly later when QCD phase transition has completed, at
$t\sim 10^{-4}\,sec$ or $T\sim 100\, MeV$, we have
\begin{equation}
\rho_\gamma\sim 6.4 \times 10^8 gm/cm^3,\qquad \rho_m \sim 1.0
\times 10^2 gm/cm^3.
\end{equation}

When the low-temperature bubbles fill up the space, the
neighboring two bubbles would in general be labeled by different
$\theta_{i,j}$ representing different but degenerate vacua - we
assume that there are infinite many choices of $\theta$; they are
degenerate but complete equivalent. As example,
see \cite{Hwang, Hwang0}. Note that the internal structure of
QCD is complicated enough to disfavor the non-degeneracy of the
vacuum. The domain wall is used to
separate the two regions. Three different regions would meet in a
line - which we call a vortex. We have to estimate the total
energy associated with the domain walls and the vortices -
particularly when these objects persist to live on for a "long"
time - say, $\tau \gg 10^{-4} sec$. These domain walls and
vortices are governed, in the QCD phase transition in the early
Universe, by the QCD dynamics.

We may start with a simple estimate - the expansion factor
since the QCD phase transition up to now. The present age of the
Universe is $13.7$ billion years or $13.7\times 10^9 \times 365.25
\times 24 \times 3600$ or $4.323 \times 10^{17}$ seconds. About
the first $10^9 sec$ period of the hot big bang is previously-believed
radiation-dominated. Consider the length $1.0\, fermi$ at $t\sim
10^{-5}sec$, it will be expanded by a factor of $10^7$ up to $t
\sim 10^9 sec$ (radiation-dominated) and expanded further by
another factor of $5.7\times 10^5$ until the present time - so, a
total expansion factor of $5.7\times 10^{12}$; changing a length
of $2\, fermi$ at $t\sim 10^{-5}sec$ into a distance of $1\, cm$
now. A proton presumably of $R=1\, fermi$ at $t\sim 10^{-5} sec$
should be more or less of the same size now; or, the bag constant
or the energy associated with the false vacuum should remain the
same.

What would happen to the pasted or patched domain walls as formed
during the cosmological QCD phase transition? According to
\cite{Hwang, Hwang0}, we note that the
solutions in previously two different true-vacuum regions cannot
be matched naturally - unless the K values match accidently. But
it is clear that the system cannot be stretched or over-stretched
by such enormous factor, $10^{12}$ or $10^{13}$. We believe that
the field $\phi$, being effective, cannot be lonely; that is,
there are higher-order interactions such as
\begin{equation}
c_0\phi G_\mu^a G^{\mu,a},\quad c_1 \phi GGG,\, ...,\qquad d_0 \phi
{\bar \psi}\psi,
\end{equation}
some maybe being absent because of the nature of $\phi$. In other
words, we may believe that the strong interactions are primarily
responsible for the phase transition in question, such that the
effective field $\phi$ couples to the gluon and quark fields; the
details of the coupling are subject to further investigations.

That is, when the field $\phi$ responsible for the pasted or
patched domain walls is {\it effective} - the $\phi$ field
couples, in the higher-order (and thus weaker) sense, to the gluon
and quark fields. It is very difficult to estimate what time is
needed for pasted domain walls to disappear, if there are no
nontrivial topology involved. If there is some sort of nontrivial
topology present, there should left some kind of topological
domain nugget - however, energy conservation should tell us that
it cannot be expanded by too many orders.

In other words, the energy associated with the cosmological QCD
phase transition, mainly the vacuum energy associated with the
false vacuum, disappeared in several ways, viz.: (1) the bag
energies associated with the baryons and all the other
color-singlet objects, (2) the energies with all kinds of
topological domain nuggets or other topological objects, and (3)
the decay products from pasted or patched domain walls with
trivial topology.

Considering just before the critical temperature $T = T_c \approx 150\,
MeV$ or $t \approx 3.30 \times 10^{-5} sec$, we have
\begin{equation}
\rho_{vac}=1.0163\times 10^{14} gm/cm^3,\quad \rho_\gamma
=5.88\times 10^9 gm/cm^3,\quad \rho_m=6.51\times 10^2 gm/cm^3.
\end{equation}
Here the first term is what we expect the system to release - the
so-called "latent heat"; we call it "latent energy" for obvious
reasons. The identification of the latent "heat" with the bag
constant is well-known in MIT and Coulomb bag models \cite{MIT,TDLee}.

This can be considered just before the cosmological QCD phase
transition which took place.

As time went on, the Universe expanded and the temperature cooled
further - from the critical temperature to the supercooling
temperature ($T_s \sim 0.95 \times T_c$ with the fraction 0.95 in
fact just a common estimate) and even lower, and then the
cosmological QCD phase transition was complete. When the phase
transition was complete, we should estimate how the energy
$\rho_{vac}$ is to be divided.

Let's assume that the QCD phase transition was completed at the
point $T_s$ (in fact maybe a little short after $T_s$). Let's take
$T_s=0.95\,T_c$ for simplicity. We would like to know how the
energy $\rho_{vac}$ is to be divided. First, we can estimate those
remained with the baryons and other color-singlet objects - the
lower limit is given by the estimate on the baryon number density:
$\rho_m=6.51\times 10^2 gm/cm^3$ or $3.65\times 10^{26}
GeV/c^2/cm^3$. So, in the volume $1.0\, cm^3$ or $10^{39}\, fermi^3$,
we have at least $3.65\times 10^{26}$ baryons. One baryon has the
volume energy (i.e. the bag energy or the false vacuum energy) $57
MeV/fermi^3 \times {4\over 3} \pi (1.0 fermi)^3$ (which is $238.8
MeV$). So, in the volume $1.0 cm^3$, we have at least $238.8 MeV
\times 3.65\times 10^{26}$ or $8.72 \times 10^{25} GeV$ in baryon
bag energy. Or, in different units $8.72\times 10^{25} /
(0.5609\times 10^{24})$ $gm/c^2$ or $155.5 gm/c^2$, compared to
$\rho_{vac}$ above ($\cong 10^{10}gm/cm^3$). Thus, only a
tiny fraction of $\rho_{vac}$ is to be hidden in baryons or other
color-singlet objects after the QCD phase transition in the early
Universe.

Where did the huge amount of the energy $\rho_{vac}$ go? In
the beginning of the end of the phase transition, the pasted
domain walls with the huge kinetic energies seem to be the main
story. A pasted domain wall is forming by colliding two domain
walls while eliminating the false vacuum in between. The kinetic
energies associated with the previously head-on collision become
vibration, center-of-mass motion, etc. Of course, the pasted
domain walls would evolve much further such as through the
decaying interactions given earlier or forming the "permanent"
structures. In any case, the total energy involved is known
reasonably - a large fraction of $\rho_{vac}$, much larger than
the radiation $\rho_\gamma$ (with $\rho_m$ negligible at this
point).

Shown in Fig. 1 is the key result of the previous papers
\cite{Hwang, Hwang0} - we wish to use it to explain the important
result. At $t\sim 3.30\times
10^{-5}\, sec$, where did the latent energy $10^{14}gm/cm^3$
evolve into? We should know that the curve for $\rho_\gamma$, for
massless relativistical particles, is the steepest in slope. The
other curve for $\rho_m$ is the other limit for matter (which
$P\approx 0$). In this way, the latent energy is connected
naturally with, or partially connected with, the curve for
$\rho_{DM}$ - in fact, there seems to be no other choice.

\begin{figure}[h]
\centering
\includegraphics[width=4in]{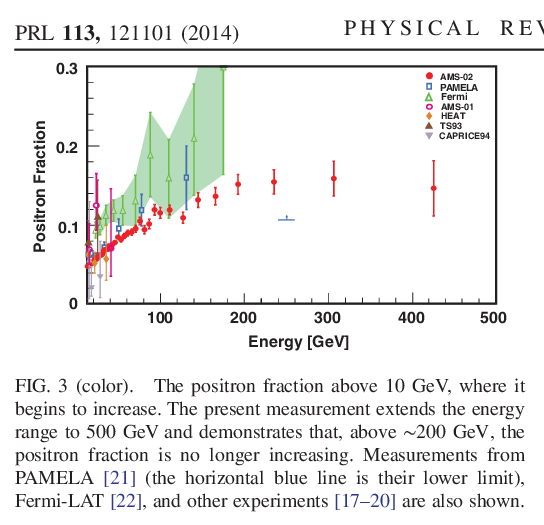}
\caption{The various densities of our universe versus time.}
\end{figure}

Coming back to Eq. (23), we could assume for simplicity
that when the cosmological QCD just took place the system follows
with the relativistic pace (i.e. $P=\rho/3$) but when the system
over-stretched enough and had evolved long enough it was diluted
enough and became non-relativistic (i.e. $P\approx 0$). It so
happens that in both cases the density to the governing equation,
Eq. (23), always looks like $\rho \propto t^{-2}$, although it is
$R \propto t^{1\over 2}$ followed by $R \propto t^{2\over 3}$.

It is so accidental that what we call "the radiation-dominated
universe" is in fact dominated by the latent energy from the
cosmological QCD phase transition in the form of "pasted" or
"patched" domain walls and the various evolved objects. In our
case, the transition into the "matter-dominated universe", which
happened at a time slightly different from $t\sim 10^9 sec$,
occurred when all the evolutions of the pasted domain walls ceased
or stopped. In other words, it is NOT the transition into the
"matter-dominated universe", as we used to think of.

In fact, the old way of thinking of the "dark matter", or the
significant portion of it, turns out to be very natural.
Otherwise, where did the $25 \%$ content of our universe come
from? Of course, one could argue about the large amount of the
cosmological QCD phase transition,
in terms of the largeness of the latent energy. The curves in
Fig. 1 might make a lot of sense \cite{Hwang,HwangWYP}.

Of course, neutrino halos could naturally be identified with the large
portion of the $25\%$ dark matter \cite{Hwang9, Hwang8}. The latent
heat (energy), on the other hand, should have some place to go - if
it does not belong to the $25\%$ dark matter, it has to go somewhere
visually, i.e., in the visual ordinary-matter way.

The role of the latent heat(energy) in the cosmological phase transitions
poses a fundamental question that deserves our further thoughts.

\medskip

\section{The Outlook}

In this paper, we use the Standard Model
\cite{Hwang417} to analyze possible phase
transition(s). For what is called "electroweak" 
phase transition, it {\it does not exist}.
The setting-in of the spontaneous symmetry 
breaking ($SSB$) is equivalent to the change
of the purely imaginary temperature. Besides 
that, it is a completely dimensionless theory
-- there is no way to maneuver on the theory.

We also argue that from the macroscopic scale ($1\,cm$) to the microscopic
scale ($10^{-8}\,cm$) the variables get reduced (elimination from momenta/coordinates
to coordinates) by the quantum principle and so, at the level of the distance
$10^{-20}\, cm$ or so, further reduction comes from changing commutative coordinates
to noncommutative coordinates when the quantum era ($10^{-8}\,cm$) has changed to the
super-quantum era ($10^{-20}$\,cm or small). In fact, we realize that H. Snyder
\cite{Snyder}, in as early as 1947, already tried to extend the space-time to
a discrete one while observing Lorentz symmetry - it provides a nice
formulation of our "super-quantum" idea. The noncommutative idea comes from
\cite{Snyder, Connes}, which may have a natural place to enter in our
thinking.

We also have a few remarks on an approximate way to treat the cosmological
QCD phase transition, a result that indicates the basic importance
of the latent heat (energy) if the phase transition would be first-order in
nature. In view of the complicated nature of the subject, we suspect that it
will remain the unsettled case for a while for many years or decades
to come. In the case of the "mass" phase transition, it is
important realize that there is no real phase transition,
according to the Standard Model \cite{Hwang417}.

We hope that our investigations would shed light on the fundamental
questions regarding the phase transitions in our Universe, and in
particular at the level of the smallest units of matter.

\bigskip

\section*{Acknowledgments}
The Taiwan CosPA project is funded by the Ministry of Education
(89-N-FA01-1-0 up to 89-N-FA01-1-5) and the National Science
Council (NSC 95-2752-M-002-007-PAE). This research is also
supported in part as another National Science Council project (NSC
99-2112-M-002-009-MY3).

\end{document}